\pdfoutput=1
\documentclass{jdfsl}

\usepackage{makeidx}  
\usepackage{upgreek}

\usepackage{url}

\usepackage{amsmath}

\usepackage[T1]{fontenc}

\usepackage{array}

\usepackage{graphicx}

\usepackage{subfig}
\usepackage{multirow}
\usepackage{array}
\usepackage{amssymb}

\usepackage{natbib}

\usepackage{enumitem}

\graphicspath{{./pdf/}{./jpeg/}{./images/}}
   \DeclareGraphicsExtensions{.pdf,.jpeg,.jpg,.png}

\begin{document}

\title{Leveraging Decentralization to Extend the Digital Evidence Acquisition Window:\protect\\Case Study on BitTorrent Sync}

\author{Mark Scanlon \and Jason Farina \and Nhien An Le Khac \and Tahar Kechadi}


\institute{School of Computer Science \& Informatics,\\
University College Dublin,\\
Belfield, Dublin 4, Ireland.\\
\{{mark.scanlon,an.lekhac,tahar.kechadi\}@ucd.ie, jason.farina@ucdconnect.ie}
}

\abstract{
File synchronization services such as Dropbox, Google Drive, Microsoft OneDrive, Apple iCloud, etc., are becoming increasingly popular in today's always-connected world. A popular alternative to the aforementioned services is BitTorrent Sync. This is a decentralized/cloudless file synchronization service and is gaining significant popularity among Internet users with privacy concerns over where their data is stored and who has the ability to access it. The focus of this paper is the remote recovery of digital evidence pertaining to files identified as being accessed or stored on a suspect's computer or mobile device. A methodology for the identification, investigation, recovery and verification of such remote digital evidence is outlined. Finally, a proof-of-concept remote evidence recovery from BitTorrent Sync shared folder highlighting a number of potential scenarios for the recovery and verification of such evidence.
}


\maketitle

\section{Introduction}

Cloud-based file synchronization services have become very popular in recent years offering a remote backup of data paired with the automation of data across multiple devices. As an indication of the growing popularity of these tools: the largest of these providers, Dropbox, announced serving over 275 million users in April 2014, up from 200 million users in November 2013 \cite{dropbox}. Highlighted by the recent privacy concerns raised by the revelation of the extent and reach of the National Security Agency (NSA) in the United States and their global Internet monitoring system, PRISM, many corporate and home Internet users have recently taken a keen interest in the active protection of their privacy and of their data. \citet{duranti2013records} conducted a cloud security survey with corporate information storage influencers (such as Records Managers, Information Officers, Archivists, etc.). The authors found found that 56\% of the 323 respondents were against cloud storage adoption due to the potential security risk, and 40\% of respondents had concerns over the privacy risks.

BitTorrent Sync (BTSync) was launched as an open alpha release in April 2013 to provide users with a private, secure, decentralized file synchronization tool capable of providing similar functionality to its cloud-based alternatives. The most recent usage figures released by BitTorrent Inc. are from December 2013 claiming over two million active users -- doubling from the previous month. From a privacy standpoint, decentralized file synchronization solutions are more desirable to their cloud-based counterparts as the only method available for any unauthorized user/body to access any private data would require law enforcement to seek a warrant to physically investigate the user's devices (in most jurisdictions). Whether collecting evidence from a cloud-based or cloudless solution, recovery and verification of remotely gathered digital evidence poses a difficult task to digital investigators in terms of technical best practice and legal responsibilities.

The need for a solution to this remote digital evidence retrieval problem is compounded further by the prevalence of smartphones and tablets in today's world. These devices typically have much smaller local storage capacities when compared with their desktop/laptop counterparts. As a result, the need for consumer technology users to have synchronized cloud/remote data storage is ever increasing. 

When configuring a newly purchased mobile device based on Android, iOS or Windows operating systems, it is encouraged to enable cloud synchronization options for the user's mobile device information, such as contacts, emails, calendar events, photos, documents, etc. Accessing data stored on cloud-based storage services is a seamless process to the user for any mobile device with a data connection.

When remotely stored information is accessed on a mobile device, the data is downloaded to a local temporary cache while the user is viewing it and generally is deleted when the user closes the mobile application or restarts the device. This common feature of mobile applications presents a difficult task for the digital forensic investigator to identify, access and analyse this potentially rich digital evidence source stored remotely in the cloud and/or on synchronized remote devices.\\

\subsection{Contribution of this Work}
\label{contribution}
While some work has been conducted on the recovery of evidence from file synchronization services, it mainly focuses on the gathering of files and logs stored on the local machine's hard drive. Recovering evidence from cloud-based solutions is generally conducted through a browser interface or client application synchronization, requiring the authentication details for the service. At the time of writing, the authors were unable to identify any publications focused on an ``after the fact'' recovery of locally compromised or unrecoverable evidence from a decentralized file synchronization service. 
The contribution of this work can be summarized as follows:
\begin{itemize}
\item This paper presents a methodology for the forensically sound remote recovery and verification of digital evidence from decentralized file synchronization services. This can enable forensic investigators to overcome a number of counter-forensic techniques potentially employed by cybercriminals to cover their tracks.

\item A proof-of-concept implementation of the methodology is described for BTSync, outlining the entry points to the investigation, the network specific knowledge required for the remote recovery of digital evidence and methods available to the digital investigator for the verification of any remotely collected evidence.\\
\end{itemize}

\section{Related Work}
\label{related}

The following subsections outline work related to that presented in this paper in the areas of remote evidence acquisition, cloud-based file synchronization forensics, mobile application forensics and the court admissibility of remotely gathered evidence. An introduction to the BTSync application and its behavior is also included.\\

\subsection{Remote Evidence Recovery}

A client server based system for remotely gathering forensically sound disk images over the Internet is outlined in an article by \citet{scanlon2010online}. This system was based on a live forensics scenario whereby the suspect machine is booted using a Linux based live CD or USB key in order to take a verifiable, remote clone of any storage device on the machine. The evidence gathering and verification process utilized frequent SHA512 hashing of ``chunks'' of the remote hard drive in combination with a hash of the remote volume in its entirety. The authors found that the division of large evidence examples, its transmission over an encrypted Internet connection to a server in a forensic laboratory and the subsequent recombination process did not interfere with the resultant hash values when compared against those of the original volume.\\

\subsection{Cloud-based File Synchronization Forensics}
\label{cloudforensics}
The task of performing evidence acquisition from cloud-based file synchronization systems is the most analogous to the work presented as part of this paper, despite using a centralized server. \citet{chung2012digital} proposed a novel process model for the investigation of cloud storage services outlining best practices for forensic investigators. Research conducted into the evidence recovery from cloud-based file synchronization tools splits into two evidence gathering tactics: local cloud evidence acquisition and remote cloud evidence acquisition. 

Local cloud evidence recovery focuses on recovering cloud storage remnants through hard drive analysis. In a volume of work conducted by \citet{quick2012masters}, the local remnants of deleted files is analyzed across Dropbox, Google Drive and SkyDrive (now OneDrive). The metadata that remained was sufficient to prove that a cloud-based file was present on the local drive after deletion. The authors also proved that the act of downloading the data from the remote location using a browser or synchronizing using the client application does not change the hash of the file or any associated cloud metadata \citep{Quick2013266}. The only metadata that was different on the local machine when compared to its cloud stored counterpart was the file creation/modification dates.

Accessing remote digital evidence stored on the servers of these cloud file synchronization service providers is an arduous task for digital forensic investigators. In 2014, Federici described a tool built for the collection of evidence from the cloud called Cloud Data Imager (CDI) built as an extension to the work conducted on local cloud related remnants by \citet{Federici201430}. CDI facilitates the read-only access to remote evidence stored on Dropbox, Google Drive and SkyDrive. This tool relies on the recovery of the cloud service's username and password or an access token string from the local machine for remote authentication.

\subsection{Recovering Evidence from Mobile Applications}
\label{recovering}

There are three kinds of mobile device forensic acquisition: logical, physical and mechanical disassembly of the device. For the purposes of this paper, first two types are of interest. Physical acquisition is used to directly clone all data from the mobile device's storage. It normally requires on ``jailbreaking'' or ``rooting'' the mobile device and using SSH communication to access the device's storage \citep{zdziarski2008iphone}. Logical acquisition requires the availability of system backups of the relevant device stored the suspect's computer. The drawback for this method is that it cannot extract deleted data or accessing the system partition as these backups generally restricts evidence collection to the device's user data partition \citep{hoog2011iphone}. With regards to iOS, unallocated space has been largely inaccessible since iOS 4.0. This is due to the encryption approach used by Apple to prevent deleted files from being recovered. A new encryption key is created for each file living on the file system and when the file is deleted, that key is removed resulting in an unrecoverable file.

\citet{grispos2013smartphones} conducted a comprehensive analysis of Dropbox's, Box's and SugarSync's mobile cloud synchronization applications for Android and iOS focusing on the recoverability of cloud synchronized data. The authors discovered that any files explicitly marked for offline access were recoverable from both mobile operating systems (even if the file was deleted from the cloud service). However, files that were merely viewed/accessed on the mobile devices were generally not recoverable, although some associated metadata remained.\\

\begin{figure*}[!t]
\centering
\includegraphics[width=1\textwidth]{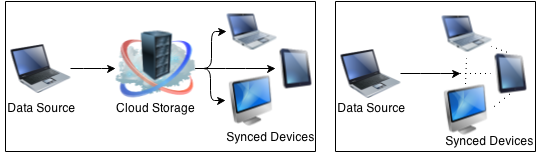}
\caption{Example Data Flow Highlighting the Difference Between Cloud-based (Left) and Decentralized (Right) Synchronization}
\label{fig:dataflow}
\end{figure*}

\subsection{Court Admissibility of Remotely Gathered Evidence}
\label{admissibility}

As with any digital forensic evidence, the court admissibility of the gathered information is largely reliant on the experience of the expert witness and comprehensive documentation of the process. In many nations, the law lags behind in formally recognizing the development of new forensic processes. \cite{kenneally2005confluence} published a paper outlining the legal concerns in the United States surrounding the live acquisition of remote digital evidence. Kenneally surmises that the admissibility of any digital evidence is reliant on the integrity of the individual forensic practitioner and that the acceptance of any new evidence acquisition methodology requires endorsement from law enforcement, forensic investigators and authoritative bodies. Baring this status quo in mind, any evidence gathered using the methodology outlined in the following section should be reliable and reproducible provided sufficient documentation of the process used accompanies any evidence presented in a courtroom. This view is echoed in many other forensic publications, such as the view of Federici who states that a digital investigator is not bound to a specific set of tool or approaches, so long as he can justify his actions \citep{Federici201430}. In order to investigate a suspect's mobile device, generally a warrant is required. In a corporate investigation, a warrant is not needed to investigate any equipment owned by the company. However, it is illegal to seize or to investigate a personal mobile phone without the consent of the owner of the device \citep{zdziarski2008iphone}.

\subsection{BitTorrent Sync}

BTSync shared folders rely on secrets to synchronize data between peers. Secrets are the unique identifiers used by BTSync to differentiate between all shared folders across the network. In order for the 33-byte secrets to be human readable, they are displayed to the user using Base32 encoding. BTSync facilitates the generation of three categories of secrets for the synchronizing of data contained within specific folders - master secret (read/write), read-only secret and an encrypted secret. Any client application that is supplied this secret will synchronize all files with any remote machines with the same secret.

BTSync, as a decentralized solution differs to the cloud based services outlined above (as can be seen in Figure~\ref{fig:dataflow}), while offering much of the same synchronization functionality to the end user. The key difference is that any data transfer using BTSync can only occur if at least one synchronized device is online. This makes the recovery of the data quite different from a digital forensic perspective due to there being no centralized file system, redundant data block algorithms or requirement of cooperation from a cloud storage provider to perform the investigation. Due to the fact that BTSync uses a distributed hash table (DHT) to disseminate peer information, there is also no central authority to manage authentication or log data access/modification attempts. A suspect file identified on a system may have been downloaded from one or many sources and may have been subsequently uploaded to one or many recipients. 

The pertinent metadata required for the remote recovery of data are \citep{Farina2014S77}:

\texttt{33-byte Secret} -- This is the unique identifier of any given shared folder on the BTSync network. It consists of a single byte indicating the access level followed by a 32-byte application generated salted hash of the folder.

\begin{figure}[!h]
\centering
\includegraphics[width=0.48\textwidth]{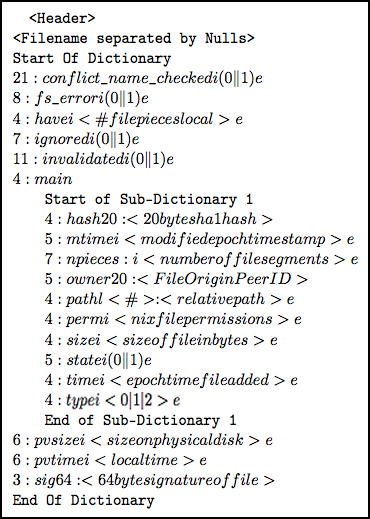}
\caption{Bencoded File Information}
\label{fig:fileentry}
\end{figure}

\texttt{<ShareID>.db} -- This file is a SQLite3 database. The database describes the contents of the shared directory corresponding to the ShareID. It contains two tables; \texttt{files} and \texttt{meta}. The files table contains filenames, paths, file sizes and SHA1 hash values for each individual file. The meta table contains information pertaining to the file (filename, file size, SHA1 hash) and data relating to each of the pieces the file is split up into for synchronization purposes (number of pieces, size of each piece, piece SHA1 hash value and a SHA1 hash value of all concatenated piece hash values). Sample bencoded data contained in the \texttt{<ShareID>.db} file is similar to that outlined in Figure~\ref{fig:fileentry}. Note bespoke file information is included in angle brackets, ``\texttt{\#}'' represents the variable length of the associated value).

In 2014, Scanlon et. al published a network investigation methodology for BTSync \cite{scanlon2014methodology}. The paper performs an analysis of the network traffic that is generated during regular BTSync operation. The detailed description of LAN multi-cast, tracker, relay server, known host and file synchronization communication are outlined. Several scenarios for the potential malicious use of the network are described including industrial espionage, cloudless backup of illicit files, encrypted remote P2P backup, dead drop, secure messaging, piracy, server-less website hosting and malicious software distribution. Any of these scenarios could make the remote recovery of digital evidence from the network relevant for investigation.\\

\begin{figure*}[!t]
\centering
\includegraphics[width=1\textwidth]{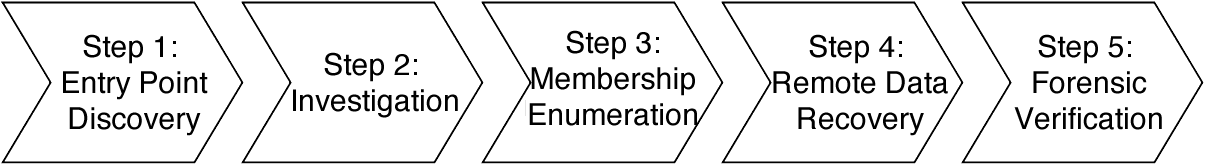}
\caption{Steps Involved in Remote Evidence Acquisition}
\label{fig:steps}
\end{figure*}


\section{Remote Evidence Acquisition Methodology}
The principle difference between evidence acquisition from a centralized (generally cloud-based) system and a decentralized (cloudless/serverless) system is the reliance on the nodes themselves to maintain and update the synchronized data.
Example scenarios where the acquisition of the remote synchronized evidence may be necessary to the investigation include:

\begin{itemize}
\item \textbf{Deleted local data (unrecoverable)} -- If the locally stored copy of the data has been securely deleted or corrupted, then the only available forensically sound source to recover that data may be from the remotely synchronized machines.
\item \textbf{Identification of remote machines after local uninstall} -- The complete uninstallation of the application may be employed as a counter-forensic technique in order to obfuscate the participation in any remote replication. Using recovered remnants of the installation can lead to the identification of remote IP addresses.\\\\
\item \textbf{Identification of remote machines sharing the same file} -- Calculation of the SHA1 hash of any incriminating evidence could aid in the identification of remote IP address sharing the same file.
\item \textbf{Data modified offline} -- If synchronized data has been modified on the local machine while offline and the changes made are pertinent to the investigation, remote synchronization can aid in identifying what changes were made.
\item \textbf{Evidence accessed on a mobile device} -- In this scenario, the entry point to the investigation may be evidence gathered of a suspect viewing or accessing the data on a mobile device. This data may only ever be stored on the mobile device in a temporary cache.
\end{itemize}

The remote acquisition of evidence identified to have been accessed (or previously stored) on the local machine involves a five step process, as can be seen in Figure \ref{fig:steps}. Each of these steps is outlined in greater detail the following subsections.\\

\subsection{Discovery -- Identifying Entry Points}
Before an investigation can begin the investigator must identify what possible entry points exist for that particular protocol. These entry points will identify resources that can be used to create an accurate profile of the data being replicated to or from the suspect device. In a cloudless replication system there may not be a single online accessible repository of the entire dataset. The investigator must therefore not only identify what data was potentially present on the suspect system but also what data was removed and whether it was replicated from or to another system that is still accessible. For this reason the investigator may be required to retrieve data from a variety of sources such as network traffic, RAM analysis, local system analysis (hard drive forensics), mobile device forensics performed on any network capable device that the suspect user may have access to. Often more than a single resource will be required to build a complete enough picture to enable accurate retrieval. The most common entry points to an investigation are described in more detail below.\\

\subsubsection{Network Traffic Analysis}
\label{NDIS}
Network traffic analysis requires the investigator to examine the logs and raw data of the communication between replicating peers in an effort to determine what data was shared and in which direction the data was replicated. Most often an investigator will initially just have log files to work from that provide high level details of the network traffic. The investigator should therefore be familiar with the traffic signature of the protocol being investigated. In the case of an on-going investigation where a traffic source is suspected, the investigator may have access to network capture files which will provide a much clearer and in depth image of the nature of the communication. Again, the investigator should be familiar with the traffic signatures in order to correctly configure a capture or display filter to make analysis faster and clearer. This will also reduce the risk of inadvertently infringing on the privacy of any non-suspect users.

\subsubsection{File System Analysis}
\label{LSDIS}
-- A second entry point results from the analysis of the suspect workstation. In this scenario, the investigator has become aware that this system is participating in a file synchronization network. Once the investigator knows of the presence of a file replication network in use from a suspect system, the decision can be made whether to perform a live forensic investigation or whether to image the system and perform postmortem analysis. Depending on the type of system involved the investigator will need to be familiar with the behavior of the protocol to best determine the advantages and disadvantages of each option. \\

\subsubsection{Memory Analysis}
\label{RAMDIS}
-- Memory analysis is another entry point available to the investigator and can prove vital if the replication protocol allows any form of encryption or file security. In order to process the files any application will usually load potentially required security keys and secrets into the process address space in RAM for quick access and temporary storage. Once the system is powered off or over a period of activity any data stored in RAM will be lost or overwritten (such as traces of connections to remote ports or application communications received). If the application is still installed and running however, there will still be a base level of information stored in RAM that may provide evidence or at least may verify or help explain items of interest found on other entry points. \\

\subsubsection{Mobile Device Analysis}
\label{mobileentry}
-- When a mobile device is mounted as a file system, its raw partition looks like one large file that contains both live and deleted data. Data-carving tools such as Scalpel\footnote{https://github.com/sleuthkit/scalpel}, Ontrack EasyRecovery\footnote{http://www.krollontrack.com/data-recovery/recovery-software/}, etc., can scan the disk image for deleted files. The investigation should focus on the recovery of files relating to the application, e.g., log files, metadata files, synchronized files from the temporary cache, etc. These files should provide an entry point into the shared information or aid in the verification of entry point information gathered from other sources.\\

\subsection{Investigation -- Uncovering Local Metadata}
Once the entry points have been identified the investigator must assess the priority of their analysis. In general, this priority will follow the same order as that laid out in the \citet{acpo} (ACPO) guidelines taking volatility into account.  The investigator must also identify where one source of data can be used to verify the conclusions drawn from another data source. For example, network traffic captured from live traffic or gathered from network logs may be cross referenced with network data extracted from a RAM dump taken from the suspect system. After the available entry points have been discovered, the investigator must determine the reliability of the evidence source with respect to age and potential tampering. A set of relevant data evidence must then be generated and profiled in order to create a subset that contains data that has been changed, deleted or otherwise tampered with. In general this process will take the form of identifying data that:

\begin{table*}[!htpb]
\centering
    \begin{tabular}{|l|l|l|l|l|l|l|}
    \hline
    \textbf{Filename} & \textbf{Filesize} & \textbf{SHA1}  \\ \hline
badfileone.txt & 19B & \texttt{58B47FB1467AEB0BEFE6FE1BD6255A5C24B552A0} \\ \hline
    badfiletwo.txt & 124B & \texttt{B47C7586BC82B27A8441A8E4C07F77874CC67557}  \\ \hline
    badfilethree.txt  & 152B & \texttt{3598492B4D1CE5FAFD9EF76E8FA54C8F55E0716A} \\ \hline
    \end{tabular}
\caption{Files Included in Source Share}
\label{listofiles}
\end{table*}

\begin{itemize}
\item Reveals what evidence exists such as a share or file manifest, a directory listing or some log of data shared
\item Reveals what evidence has been altered or destroyed and when this occurred. Most replication systems use timestamps to record alteration times so the required replication direction can be determined to ensure later changes are not overwritten by older. 
\item Describes if and where a viable copy of the evidence can be retrieved. The investigator must be aware of any versioning facility available within the protocol and how the protocol handles local changes to replicated data.
\item Allows access to the remote copy of the evidence. This could be in the form of certificates, passwords or phrases or even just some combination of system files that are checked by the application to determine system level access rather than user or account level permissions.\\
\end{itemize}

\subsection{Enumeration --\\ Identifying Remote Data Stores}
Based on the entry point information collected, a peer discovery survey must be carried out to determine the totality of the participants in the network or share. Each potential remote synchronization node must be identified and queried to determine if recovery is possible from that source. Ideally, a source of remote evidence that has directly synchronized with the suspect's device would be identified for maintaining provenance.\\

\subsection{Recovery -- \\Downloading of Remote Evidence}
With the remote peers identified and prioritized the investigator should now determine what required version of the data is present and available. Any retrieval performed by the investigator should be carried out through programmatically ethical means without unduly alerting the remote host and in a forensically sound manner. This step is best performed either through the software used for replication originally or through a bespoke version of client software that uses the underlying protocol to retrieve the data in the same manner as it was initially gathered.\\

\subsection{Verification -- \\Ensuring Forensic Integrity}
Due to the sensitive nature of digital evidence collection, it is imperative that the data collected by any forensic tool is completely verifiable and identical to the original source. In order for any remotely gathered evidence captured using this methodology to be court admissible, it must be proven to be true to the original data that was stored/accessed on the suspect device. Each of the aforementioned synchronization services rely on frequent hashing of each file to ensure data integrity and to detect any local changes to be updated to other synced devices. Gaining access to these hashes or remnants thereof can quickly facilitate the verification of any remotely gathered evidence.\\


\section{Proof-of-Concept Example}

To confirm our methodology, we conducted a proof-of-concept investigation on the remote recovery and verification of evidence from a BTSync share. We selected BTSync due to its mostly decentralized nature. BTSync is based upon BitTorrent, a popular file transfer protocol, which has made significant strides towards complete decentralization in recent years. The BTSync application allows the user to configure most aspects of the file synchronization process and the peer discovery process. \\

\begin{figure*}[!thbp]
\centering
\includegraphics[width=0.9\textwidth]{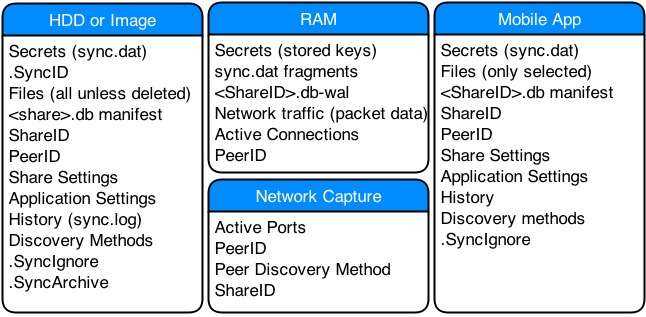}
\caption{Metadata Available from Each Entry Point}
\label{fig:POC}
\end{figure*}

\subsection{Setup}

\begin{enumerate}
\item A source machine \texttt{ComputerA} was set up with a BTSync shared folder containing the files outlined in Table~\ref{listofiles}. 
\item The file \texttt{badfileone.txt} was then deleted before any remote synchronization was allowed to take place. 
\item A second machine \texttt{ComputerB} had the BTSync read-only secret added to the application and synchronization was allowed to complete. 
\item The file \texttt{badfilethree.txt} was subsequently securely deleted from \texttt{ComputerB} to simulate the anti-forensic destruction of evidence.
\end{enumerate}

\subsection{Entry Points}
\label{POC:Entry}
The entry points possible for a BTSync investigation are:
\begin{enumerate}
\item Extracting metadata from a local memory snapshot taken while the application is running.
\item Analyzing BTSync network traffic.
\item Local file system forensics.
\item The BTSync mobile application and associated files.
\end{enumerate}

The analysis of entry points 1-3 is outlined in the two BTSync publications previously discussed \citet{Farina2014S77, scanlon2014methodology}. For the purpose of the experiment outlined in this section, a logical forensic acquisition of an iPhone 4S with a synchronized BTSync application was conducted with Oxygen Forensic Suite 2014 (as the iOS version used was iOS 7.1.1). The pertinent BTSync files in the application's folder structure are all stored in a \texttt{BitTorrent Sync} subfolder. This folder contains files such as \texttt{settings.dat}, \texttt{sync.log}, \texttt{sync.pid}, \texttt{sync.dat}, etc. The synchronized data files are stored in the \texttt{Storage} subfolder where the \texttt{.SyncID}, \texttt{.SyncIgnore} can also be found. Assuming the investigator's objective is to recover deleted files on this device, common forensic mobile device data recovery tools, such as Ontrack EasyRecovery and Oxygen Forensic Suite, are not fit for purpose. As a result, the only other avenue available to recover this data is to perform evidence recovery from remotely synchronized machines.\\

\begin{table*}[!htpb]
\centering
    \begin{tabular}{|l|c|c|c|c|c|c|}
    \hline
     \textbf{Evidence Item}     & \textbf{Network} & \textbf{RAM} & \textbf{sync.dat} & \textbf{.SyncID} & \textbf{<ShareID>.db} & \textbf{sync.log} \\ \hline
     ShareID            &R               &R     &R        &R       & ~            &R        \\ \hline
     Secret             & ~               &R      &R        & ~       & ~            & ~        \\ \hline
     PeerID             &R               &P      &          & ~       &R            &          \\ \hline
     File List          & ~               &P      &          & ~       &R            &R        \\ \hline
     File Hash          & ~               &P      & ~        & ~       &R            & ~        \\ \hline
     Remote Peers       &R               &P      & ~        & ~       &R            &R        \\ \hline
     Ports              &R              &R      & ~        & ~       & ~            &R        \\ \hline
    \end{tabular}
\caption{Location of BTSync Specific Metadata (R~=~Recoverable, P~=~Possibly Recoverable)}
\label{metadata}
\end{table*}

\subsection{Investigation}
\label{POC:Invest}
With the entry points identified, the required secrets, ShareID, log files and file hashes can be gathered. In order to verify the gathered information required for the investigation, a secondary source should be used to corroborate the data. Figure~\ref{fig:POC} outlines the list of BTSync share specific information available from each of the entry points. If two or more of these sources correspond, the higher the reliability of the gathered information. If conflicting information is gathered from two or more sources, it is likely that this inconsistency may result from a out of date synchronization or local file I/O handling. In order for the investigation to be possible, the minimum required metadata from any source is the secret.

The BTSync protocol transfers a full copy of the file manifest (contained in \texttt{<ShareID>.db}) in its current local state from the source system before any replication is initiated. Any device that is associated with a share will store a full manifest listing the file metadata and corresponding state on the source machine. This manifest can also contain metadata from each peer contributing to any modifications on the shared folder. A mobile phone that has the application installed and a secret applied may store a full file manifest, which gets updated whenever the application is run. This update will occur from any active remote machine, even if no file is ever accessed on the mobile device.

The most reliable source of evidence is that gathered from the share, log and application files stored on a computer's hard drive, e.g., data forensically discovered on \texttt{ComputerB} in the example investigation. However, the memory image can contain data that does not have the opportunity to be recorded to the hard drive, such as open ports. It may contain data that supersedes that on the physical drive such as the contents of the \texttt{<ShareID>.db-wal} (write ahead log) file. This file acts as a cache of data waiting to be written to the \texttt{<ShareID>.db} SQLite3 database. In testing, the metadata information outlined in Table~\ref{metadata} are recoverable from the marked sources.

The investigator should try to ascertain which secret extracted from RAM is associated with which shared folder. ShareIDs are derived from the secret.

In any shared folder, an investigator can find the \texttt{.SyncID} file. This file contains the ShareID that corresponds to that shared folder. The investigator must use a hex viewer to extract the ShareID from the \texttt{.SyncID} file. To determine the location of the shared folder associated with any given secret, the investigator can search in the \texttt{sync.dat} file located the directory locations outlined below:\\ \\ \footnotesize \texttt{$\sim$\textbackslash AppData\textbackslash Roaming\textbackslash BitTorrent Sync} (Windows)\\\\
 \texttt{$\sim$/Library/Application Support/BitTorrent Sync/} (Mac OS X)\\
\\\texttt{[folder where BTSync is extracted]/.sync/} (Linux)\\\\
\texttt{[Applications]/com.bittorent.BitTorrentSync/\\Documents/BitTorrent Sync/} (iOS)
\\\\
\normalsize

The \texttt{sync.dat} file is a bencoded dictionary consisting of a series of blocks describing each share active on the system and the preferences set for that particular share. The entry for each share contains the information outlined in Table~\ref{syncdat}.

\begin{table*}[!htpb]
\centering
    \begin{tabular}{|l|p{8.3cm}|}
    \hline
	\textbf{Key}     & \textbf{Value}  \\ \hline
\textbf{path}	& Path to the share folder \\ \hline
 \textbf{secret} & 33 character secret associated with that share \\ \hline
 \textbf{pub\textunderscore key} &	256-bit public key here. Only shares that have performed a handshake online will have a public key assigned.\\ \hline
 \textbf{stopped\textunderscore by\textunderscore user} & Binary toggle (0/1) if the synchronization of the share was canceled by the user\\ \hline
 \textbf{use\textunderscore dht} & Binary toggle to use DHT for peer discovery\\ \hline
 \textbf{use\textunderscore lan\textunderscore broadcast} & Binary toggle to use LAN broadcast messages\\ \hline
 \textbf{use\textunderscore relay} & Binary toggle to use a relay server if a remote peer cannot be contacted directly\\ \hline
 \textbf{use\textunderscore tracker} & Binary toggle to use  a tracker server (\texttt{t.usyncapp.com}) for peer discovery\\ \hline
 \textbf{use\textunderscore known\textunderscore hosts} & Binary toggle to use a list of host IP for direct contact\\ \hline
 \textbf{peers} & Start of the list of recorded peers that have interacted with the share\\ \hline
	 \textbf{id} & Remote PeerID\\ \hline
	 \textbf{last\textunderscore sync\textunderscore completed} & Epoch timestamp of the last time the peer was synchronized to or from\\ \hline
\end{tabular}
\caption{List of \texttt{Key:Value} Pairs for Each Share Contained in \texttt{sync.dat}}
\label{syncdat}
\end{table*}

From this file, the shared folder location can be recovered and subsequently the \texttt{.SyncID} file. Once the ShareID is known, the investigator can examine the corresponding \texttt{<ShareID>.db} file (described above) and can look for file entries that have the following \texttt{Key:Value} pairs:

\begin{itemize}[label={}]
  \item \textbf{state:i2e} -- The file was deleted on the source system and was either never synchronized or was moved to the \texttt{.SyncArchive} folder (if set in \texttt{settings.dat}) and stored for 30 days before deletion from the system.
  \item \textbf{invalidated:i1e} -- The file was locally deleted or modified and no longer receives updates from the source system as a result. The \texttt{mtime} value will store the timestamp of when the file was invalidated. For a modified file, the \texttt{hash20} value stored in the \texttt{<ShareID>.db} file is the hash value of the last known valid version of the file, i.e., matching hash values with the source file.\\
\end{itemize}

\subsection{Enumeration}
\label{POC:Enum}

Next we enumerate the members of the replicating peer group. A bespoke peer discovery application was created to identify active peers for any given ShareID using each of the regular peer discovery methods, e.g, LAN multicast, tracker and DHT. Each of the discovered \texttt{IP:Port} pairs was added to a list for later querying and file synchronization. It is important to note that each peer must ``check-in'' to the tracker and DHT every X minutes in order to be returned as active from those services (X is typically in the 10-60 minute range depending on configuration -- the default value for this on our example machines is 30 minutes). As a result, the list of active peers identified will frequently change due to network churn. It is possible to identify peers that were previously contacted from the local \texttt{sync.log} file on the suspect machine (\texttt{ComputerB} in the sample investigation). This file will contain the resolved remote hostnames, the corresponding remote PeerIDs, timestamps and specific file upload/download activity from any synchronization activity.\\

\subsection{Recovery}
\label{POC:Rec}
BTSync was installed on a monitored virtual machine for remote recovery purposes. The secret was added to the application and the file synchronization process was allowed to complete. Only IP addresses gathered during the previous step were permitted for connection, facilitated through the use of the ``Known Peers'' option in the BTSync application. Remote recovery is possible if the entire file is available from one or more remote sources. Each remote machine will store pieces of each file in the shared folder (though unlikely, it is possible that no single remote machine has the entire file). The number of pieces for each file is determined by the total size divided by the defined piece length (the default piece length is 32kb). In order for a file to be recoverable, each individual piece must be offered by at least one remote machine. If only a subset of pieces are available, partial evidence recovery is still possible (resulting in a similar recovery from a file stored on a corrupted block on a hard disk).

This recovery step can also be used to collect data that has been locally deleted from a mobile device. The BTSync mobile application facilitates the automatic upload of all photos and videos taken using the device's camera and this data will remain on remote machines irrespective of whether the media is deleted from the phone or not.\\

\subsection{Verification}
\label{POCVerification}

The underlying BitTorrent protocol used by BTSync for file transfer relies on regular hash checking. The indexing of all shared BTSync files uses systematic SHA1 hashing to know when a file is updated on a remote system. The value for the \texttt{hash} key stored for each individual file in the \texttt{ShareID.db} database is calculated using Formula 1.\\
\begin{equation}
\begin{split}
	hash key = SHA1(SHA1(Piece1)||\\SHA1(Piece2)||...||SHA1(PieceN))
\end{split}
\end{equation}
\\

Comparing this gathered file specific hash to that of the corresponding downloaded files can ensure a true copy of the original is downloaded. If only a partial download is possible, the 32kb piece hashes can be used to verify each piece of against the corresponding piece of the original file. Any downloaded piece with an incorrect hash (likely as a result of a corrupted download) can simply be discarded and re-downloaded from the same remote source or any available alternative source. 

With regards to the sample investigation, we identified that \texttt{badfilethree.txt} had been synchronized and subsequently deleted from \texttt{ComputerB}. This was discovered in the \texttt{sync.log} file and confirmed by cross-referencing against \texttt{<ShareID>.db} 
(the value of the \texttt{invalidated} tag was set to ``1'' indicating that a local unsynchronized modification had taken place). 

In order to recover the file, using a forensically sound evidence recovery machine, all peer discovery methods were disabled besides the ``Known Peers'' option, where \texttt{ComputerA}'s IP address was added. \texttt{badfilethree.txt} was subsequently downloaded from \texttt{ComputerA}. The SHA1 hash of this file was found to be \texttt{3598492B4D1CE5FAFD9EF76E8FA54C8F55E0716A} which corresponds to that of the original file, as outlined in Table~\ref{listofiles} above.\\


\section{Conclusion}
\label{conclusion}
This paper outlined a methodology for the secure, verifiable, remote recovery of digital evidence from a decentralized file synchronization network. The verification of the gathered evidence is aided by the native hash-based verification of the protocol itself. Decentralized networks must use frequent hashing to verify the integrity of the synchronized information. Accessing this information can be an invaluable source of digital fingerprinting for the forensic investigator. Any discovered hashes can be used to verify the evidence gathered from remote machines to be true copies of the original. The reliance on frequent hashing of these networks can also be beneficial for the investigator when comparing these values against a list of known incriminating hash values.\\

\subsection{Weaknesses of Current Approach}
\label{weaknesses}
One weakness of the current approach is that the remote machine can detect an unknown IP connecting to it and requesting to sync the evidence. It is possible that the remote machine might disconnect/shutdown upon detecting an unknown/suspicious IP address range. Firewall/blacklist rules, e.g., the shared \texttt{hostiles.txt} on the Gnutella network, are freely available on the web detailing the IP addresses of known law enforcement, P2P monitoring agencies and research bodies.

A second weakness of this approach occurs should the remote data be deleted/altered in the time frame between the physical gathering of a suspect's devices and the subsequent remote evidence acquisition. In this scenario, the hash of the remote files will not match any locally stored hashes or indeed the files may not be recoverable at all if no remote host of the evidence can be identified.\\

\subsection{Future Work}
\label{futurework}

BitTorrent Inc. have announced that future versions of BitTorrent Sync will facilitate selective syncing, i.e., joining the share of a particular secret, but allowing the user to decide what files to download. The company have also hinted at releasing a corporate, private version of the tool eliminating the requirement for any Internet/external access. 

The aforementioned investigative tool can be expanded upon to investigate further networks, including those designed to afford anonymity to its users. This would include the investigation and remote recovery of evidence from further decentralized networks including ownCloud\footnote{http://owncloud.org/}, Syncthing\footnote{https://github.com/calmh/syncthing} and OnionShare\footnote{https://github.com/micahflee/onionshare}. 

When a sufficient number of networks have been analyzed, a universal decentralized file synchronization forensic tool should be developed to aid in future investigations.

\newpage

\bibliographystyle{plainnat}
\begin{flushleft}
\bibliography{icdf2c}
\end{flushleft}

\end{document}